# IoT Data Quality Issues and Potential Solutions: A Literature Review

Taha Mansouri, [1], Mohammad Reza Sadeghi Moghadam, [2], Fatemeh Monshizadeh, [2], Ahad Zareravasan, [3],

[1],School of Science, Engineering and Environment, University of Salford, Greater Manchester, UK.

[2],Department of Production and Operation Management, Faculty of Management, University of Tehran, Tehran, Iran.

[3],Department of Corporate Economy, Faculty of Economics and Administration, Masaryk University, Brno, Czech Republic.

**Abstract**

*The Internet of Things (IoT) is a paradigm that connects everyday items to the Internet. In the recent decade, the IoT's spreading popularity is a promising opportunity for people and industries. IoT utilizes in a wide range of respects such as agriculture, healthcare, smart cities, and manufacturing sectors. IoT data quality is crucial in IoT real-life applications. IoT data quality dimensions and issues should be considered because we require data to make accurate and timely decisions, produce commodities, and gain insights about events, people, and the environment. It is essential to point out that we cannot reach valuable results by using poor quality data. This paper aims to develop a new category for IoT data quality. Hence, we examine existing IoT data quality dimensions and IoT data quality issues in general and specific domains and IoT data quality dimensions' categories. It is worth considering that categories in the context of IoT are not many. We developed a new category in which IoT data quality dimensions and issues are separated. Concerning this category, we can get familiar with related dimensions and issues in each category. To enhance data quality dimensions and minimize data quality issues, we suggest potential solutions using Blockchain to overcome IoT's security issues.*

## 1. Introduction

Internet of Things (IoT) is the phrase comprising all kinds of things connected to the Internet. It proposes to connect everyday items, such as watches, washing machines, vehicles, ovens, animals, plants, and the environment, to communicate with each other. The IoT paradigm began with Radio Frequency Identification (RFID). This term was mentioned by Kevin Ashton firstly in 1999 (Ashton, 2009) and is widely utilized in diverse industries like logistics, smart agriculture, smart transportation, smart medical, smart health, smart home, smart safety, and intelligent environment protection. The US National Intelligence Council (NIC) estimates that by 2025 Internet connections may remain in things that we utilize every day, such as paper documents, furniture, and food package (Atzori, Iera, & Morabito, 2010).

There are a variety of definitions for IoT from different points of view. It was represented as a group of smart things linked through Radio Frequency Identification (RFID) (Ghallab, Fahmy, & Nasr, 2020). From a connection point of view, the IoT permits individuals to be connected anywhere and anytime with everything and everyone (Côrte-Real, Ruivo, & Oliveira, 2020). From a communication perspective, IoT relates to a worldwide network of objects linked to each other based on standard communication protocols (Qin et al., 2016b). From a data point of view, related smart things will become significant data producers and consumers instead of humans (Karkouch, Mousannif, Al Moatassime, & Noel, 2016a). It has numerous challenges, such as software and algorithms, performance, architecture, data quality, and hardware (Kiruthika & Khaddaj, 2015; Shah & Yaqoob, 2016). This paper focuses on IoT's data analysis issues and, to be more specific, data quality issues are taken into consideration.

Data are the worthiest asset in the IoT because they provide insights about a given entity, phenomenon, or person are used by applications to ubiquitously offer smart services (Karkouch et al., 2016a). Moreover, data are being used to make new decisions, produce new goods, and enlarge markets as well. Many studies clarify the significance of data quality (DQ) for data mining processes and the influence of low data quality upon the reliability of the outcomes (Berti-Equille, 2007; Hipp, Güntzer, & Grimmer, 2001)). Their quality, hence, should be assured.

Data quality is defined as "the adequacy of data to the purposes of the analysis" (Merino, Caballero, Rivas, Serrano, & Piattini, 2016), "the degree to which a set of inherent characteristics fulfills the requirements" (Banerjee & Sheth, 2017) as well as "how well data meets the requirements of consumers" (Karkouch et al., 2016a). For IoT, furthermore, data quality means how appropriate the gathered data from smart things (Karkouch et al., 2016a). The diversity of the sources and the volume of data result in new difficulties in the data quality field. It is important to note that practitioners and researchers aim to assess the "fitness for use" of data sets (Sabrina Sicari, Rizzardi, Miorandi, Cappiello, & Coen-Porisini, 2016b). Data quality has become one of the significant aspects of IoT because poor decisions stem from a poor understanding of data quality. To do crucial tasks, each data user needs the utilized data to meet specific criteria. These Data Quality criteria are known as Data Quality Dimensions, such as accuracy, timeliness, completeness, and reliability (Karkouch et al., 2016a). Various literature contributions identified countless dimensions that exponentially affect the quality of data produced by the IoT, such as security vulnerability, privacy preservation, accuracy, data volume, and confidence completeness. In the literature, data quality is classified into four main categories as Intrinsic, Contextual, Representational, and Accessibility (Karkouch et al., 2016a; Lee, Strong, Kahn, & Wang, 2002).

There are some articles in IoT data quality in which they examined some data quality dimensions (Côrte-Real et al., 2020; Karkouch et al., 2016a; Karkouch, Mousannif, Al Moatassime, & Noel, 2018; Qin et al., 2016b; Sabrina Sicari, Cappiello, De Pellegrini, Miorandi, & Coen-Porisini, 2016; Zubair, Hebbar, & Simmhan, 2019). Some others proposed a categorization scheme for IoT data quality (Karkouch et al., 2016a; Lee et al., 2002; Zubair et al., 2019). It is important to note that categories in the context of IoT data quality are not many. As our contributions, we split IoT data quality dimensions, IoT data quality issues and developed a new category, and reviewed potential solutions for security issues. We target reviewed solutions to security issues, as we have a good body of knowledge there. To reach our aims, we intend to answer the following questions:

1. What are the data quality dimensions in IoT?
2. What are the data quality issues in IoT?
3. What are the potential solutions for IoT security issues?

To answer these research questions, we review the most related research in section 2, followed by the research method in section 3. Then we summarize the review results regarding the IoT data quality dimensions, categories, and issues in section 4. we propose a categorization of IoT data quality categories and issues in section 5. Potential solutions for security issues are presented in section 6, and finally, section 7 concludes the paper.

## 2. Related Research

This section purposes of comparing the previous literature in the IoT data quality context. In particular, we presented kinds of literature that identified IoT data quality dimensions and issues in section 2.1. We also presented IoT data quality categories and found potential solutions to overcome some problems with the approach proposed in this paper in section 2.2.

### 2.1. Contributions in the IoT data quality dimensions and issues

Various studies have investigated several different IoT data characteristics, and some have also identified their quality dimensions. A review of data quality in IoT provides a broad survey of the quality metrics. Data quality has been explained in various ways in the literature. Apart from the broadest utilized definition, "fitness for use", it is defined: "Data quality is data that is fit for use by data consumer. Usefulness and usability, therefore, are significant features of quality" (Strong, Lee, & Wang, 1997). Data quality examines the essential characteristics of IoT data and gives a classification of the quality dimensions. In fact, data quality dimensions provide an excellent way to assess data quality. Although some authors have defined various data quality dimensions, it is essential to point out that there is no standard definition of data quality dimensions (Baqa, Truong, Crespi, Lee, & Le Gall, 2018). Furthermore, various domain-specific data quality dimensions have been defined for multiple particular applications (Guptill & Morrison, 2013). To keep track of data quality and measure them efficiently, many characteristics as data quality dimensions have been addressed.

Metzger et al. (2012) addressed only accuracy, timeliness, and trustworthiness, and the authors proposed anomaly detection techniques to remove inaccurate and noise data to enhance data quality. Another contribution claimed that validity, accuracy, and credibility are the primary data quality dimensions in IoT (Bin Guo, Daqing Zhang, Zhu Wang, Zhiwen Yu, & Xingshe Zhou, 2013). Sabrina et al. (2016) concerned about security, privacy, and data quality. They defined data quality dimensions that can be automatically assessed: completeness, accuracy, timeliness, and source reputation. Ghallab et al. (2020) focused on IoT's data analysis issues, particularly data quality issues, including uncertainty, noise, outliers, inconsistency, and missing value. This paper proposes challenges that come from outliers. Barnaghi and Sheth (2016) identified information accuracy, validity, and credibility as the key data quality dimensions to control data sources. Klein and Lehner (2009a) have utilized five dimensions (accuracy, confidence, completeness, data volume, and timeliness) to assess sensor data streams' quality. Qin et al. (2016b) have focused on uncertainty, redundancy, ambiguity, and inconsistency as the direct dimensions. Completeness, accuracy, format, and currency were considered data quality dimensions in the IoT context (Côrte-Real et al., 2020). Togneri et al. (2019) addressed data availability and veracity as data quality issues in the context of IoT. They divided data quality issues into availability (error and interruption) and veracity (unbalanced and non-correspondence of different granularity data) problems. Farooghi et al. (2018) identified IoT data quality issues, including timeliness, accuracy, completeness, usability, trustworthiness, confidence, consistency, and readability. In a paper (Sabrina Sicari, Cappiello, et al., 2016), the authors presented accuracy, completeness, and timeliness as main issues in the context of IoT. Accuracy, completeness, usability, trustworthiness, consistency, readability, accessibility, and redundancy as IoT data quality issues by authors in Firmani et al. (2016). Barnaghi et al. (2015) mentioned precision, accuracy, and granularity as issues that stem from data characteristics. Concerning contributions in the context of IoT data quality, it is possible to state that, first of all, researchers just focused on some of the data quality dimensions for instance only accuracy and trustworthiness. Although there are various literatures which concentrated on data quality dimensions, a few of them separated data quality dimensions and issues. Secondly, they ignore gathering all data quality dimensions in both general and specific domains. While we gathered data quality dimensions and issues from literatures.

### 2.2. Main contributions in the IoT data quality categories and suggesting potential solutions

Several literature contributions identified data quality dimensions which some of which were classified into four main categories. It is worth considering that data quality categories in the context of IoT are not many.

Karkouch et al. (2016a) addressed quality issues, RFID data, and data streams. They presented accuracy, confidence, completeness, data volume, ease of access, access security, interpretability, and timeliness as the primary data quality dimensions. They also considered additional data quality dimensions such as duplication (e-health and smart grids domain) and availability (e-health domain) as the IoT domain-specific. They also identified four main categories of data quality dimensions: Intrinsic, Contextual, Representational, and Accessibility. In this article, the authors reviewed data quality in IoT and introduced generic and specific data quality dimensions. Moreover, they investigated IoT-related issues affecting data quality, such as data outliers, duplication, and data leakage. Although this contribution studied IoT data quality dimensions and issues in-depth and recognized four main categories, they did not separate related data quality dimensions and issues in category. Moreover, they identified factors that put at risk the quality of data, related data quality issues, and providing solutions to overcome them were ignored. Zubair et al. (2019) have assembled a taxonomy in seven categories of the most related data quality dimensions. They identified factors physical and environmental factors that affect IoT data quality. Even though they provided a new taxonomy that includes 21 data quality dimensions, they neglect separating dimensions and issues and suggesting solutions to overcome issues. Lee et al. (2002) and strong et al. (1997) summarized data quality dimensions into four main categories, and splitting dimensions and issues in the related category did not concern. All four categories are gathered in table 1.

Table 1: Data quality categories and dimensions

| Data Quality Categories | Data Quality Dimensions | References |
|---|---|---|
| Intrinsic | Accuracy, Reputation | (Karkouch et al., 2016a) |
| Contextual | Timelines, Completeness, Data volume | |
| Representational | Interpretability, Ease of understanding | |
| Accessibility | Accessibility, Access Security | |
| Intrinsic | Accuracy, Objectivity, Believability, Reputation | (Lee et al., 2002; Strong et al., 1997) |
| Accessibility | Accessibility, Access security | |
| Contextual | Relevancy, Value-added, Timeliness, Completeness, Amount of data | |
| Representational | Interpretability, Ease of understanding, Concise representation, Consistent representation | |
| Inaccuracy | Noise, Errors, Variable Precision | (Zubair et al., 2019) |
| Timeliness | Freshness, Usability | |
| Completeness | Incompleteness, Redundancy | |
| Ambiguity | Relevance, Lack of Context | |
| Inconsistency | Data values, Metadata values | |
| Credibility | Trustworthiness, Security, Privacy | |
| Uncertainty | ----- | |

We examine existing categories of data quality dimensions in literature and their definitions.

- Intrinsic: This category has to do with the innate quality in data or data inherited, such as accuracy.
- Contextual: Dimensions in this category describe the quality of tasks utilizing data like timeliness and completeness.
- Accessibility: This addresses the accessibility of data for data users.
- Representational: Dimensions describe how data formats are comprehensible and representative, such as ease of understanding.

Analyzing mentioned studies, it can be concluded that there are three exiting gaps. First of all, Existing articles do not have a broad view of IoT data quality. In fact, they did not consider all data quality dimensions and issues. They only have focused on some of them. We gathered data quality dimensions and problems with their definitions. Secondly, it should be noted that there are some IoT data quality categories in which various dimensions were identified. However, none of them separated related data quality dimensions and issues. Developing a new category into five main categories and separating related dimensions and issues in each category is another contribution of this research. Finally, we analyzed problems and suggest potential solutions for critical ones. There is no contribution in proposing a new category to the best of our knowledge, separating data quality dimensions and issues, and suggesting potential solutions.

## 3. Research Method

Data quality is one of the critical studies in IoT and attracted much researcher attentions. An astonishing variety of articles have been published on IoT data quality. This survey carried out a systematic literature review (SLR) of the empirical studies focusing on DQ in IoT. We identified IoT data quality dimensions and issues up to December 2020. IoT data quality categories were recognized as well.

Initially, various research databases such as Google Scholar, Web of Science, Scopus, Springer, IEEE Xplore, and Science Direct are used to search 300 articles. To identify IoT data quality dimensions and issues, words such as "Internet of Things", "Data Quality", "Data Quality Dimensions", "Quality Assurance", "IoT data quality", "IoT data quality dimensions" "IoT data quality issues" and "IoT data quality dimensions" used as keywords to search published papers. The number of articles is declined to number 180 after reading the title and abstract. Finally, 105 articles remained to examine and gather existing data quality dimensions and issues and their definitions.

It should be noted that categories in the context of IoT are not many. We found five papers in which the category of IoT data quality have been developed and analyzed. Then, to find potential solutions to overcome security problems, keywords like "IoT solutions", "IoT security challenges", "IoT Security", "IoT Privacy", "IoT Artificial Intelligence", and "IoT and MachineLearning", were explored in research databases mentioned above. Reviewing the articles which contained mentioned keywords, potential solutions were extracted. Addressing issues like privacy and security have already been covered. Apart from reviewing security and privacy issues, we suggest new potential solutions to overcome problems like trustworthiness and timeliness. The research steps are illustrated in Figure 1.

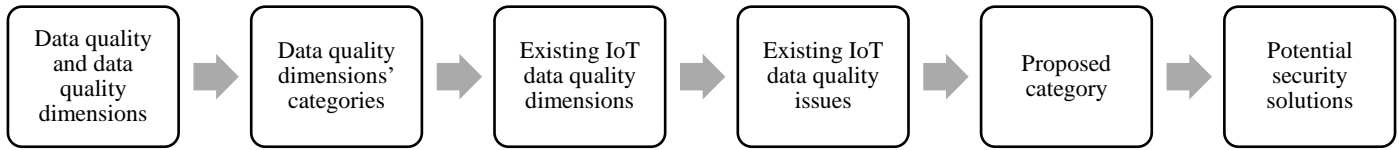

Figure 1: Roadmap of the article

## 4. Findings

Numerous features are usually correlated with data in the context of the IoT. Whereas some of these features might be viewed as omnipresent, other features are not general and significantly depend on the context. Below, we have gathered data quality dimensions and issues, along with their definitions. It is worth considering that there is no specific definition for each dimension and problem, and each of them has various definitions.

### 4.1. IoT data quality dimensions

**Accuracy** measures the closeness of captured values of data points with their original values; in other words, accuracy has a close connection with correctness (Anagnostopoulos & Kolomvatsos, 2016; Ramaswamy, Lawson, & Gogineni, 2013b; Sabrina Sicari, Cappiello, et al., 2016; Wei et al., 2019b). **Usability** regards the amount of time that data can be maintained before it comes devoid of value (Zubair et al., 2019). **Relevance** registers how much of the obtained data is of worth to an application. It may cause ambiguity if the application takes data with no utility or is not demanding (Zubair et al., 2019). **Privacy** assures limits on a person who is permitted to access the data (Zubair et al., 2019). **Currency** is the user's perception of the degree of data that is up to date (Setia, Setia, Venkatesh, & Joglekar, 2013). **Timeliness**, latency, currency, or volatility display freshness of captured data and their punctuality regarding the application context (Anagnostopoulos & Kolomvatsos, 2019; Ramaswamy et al., 2013b; Sabrina Sicari, Rizzardi, et al., 2016b; Wei et al., 2019b). **Completeness** defines the extends to which values are available in a dataset (Wei et al., 2019a). Moreover, this measure has a close relationship with non-imputed values in a data stream according to the application requirements (Anagnostopoulos & Kolomvatsos, 2019). **Trustworthiness** is associated with the concept of source reputation and reliability and defines whether the sensor feed was collected and processed by genuine infrastructures (Ramaswamy, Lawson, & Gogineni, 2013a). **Availability** refers to the amount of time a sensor feed is operational and available for use (Mavrogiorgou et al., 2019b; Ramaswamy et al., 2013b). **Validity** or freshness means how much is the reading value still valid in the original sensor during reporting (Ramaswamy et al., 2013a). **Security** is the attempt to protect the privacy, confidentiality, and integrity of sensor data (Karkouch, Mousannif, Al Moatassime, & Noel, 2016b). **Confidence** demonstrates the maximum probability of the expected statistical error to occur (Anagnostopoulos & Kolomvatsos, 2016; Karkouch et al., 2016a). **Volume (Throughput)** refers to the amount of raw data that is expected to be processed to reach the target information (Anagnostopoulos & Kolomvatsos, 2016; Karkouch et al., 2016a). **Ease of access** shows how easy is the process of data retrieval (Karkouch et al., 2016a). **Interpretability** means data has a meaningful and easy to interpret schema (Karkouch et al., 2016a). The **granularity** is a spatial and temporal dimension that measures how detailed are the stored data. Although different applications have different requirements for the level of granularity, it directly affects some other qualitative dimensions such as timeliness and completeness. Moreover, using some interpolation models such as linear, polynomial interpolation, and Gaussian is beneficial to reducing data granularity. While more detailed data is requested, the source can be used, and while less detailed is needed, the application would access to sampled or aggregated data (Barnaghi, Bermudez-Edo, & Tönjes, 2015). **Capacity** is related to concurrent access and shows the maximum extent of concurrency (Wei et al., 2019b). **Format** refers to "the user's perception of how well the information is presented". **Frequency** (temporal resolution) shows the reading rate regarding a specific time, such as reading among 2 minutes (Ramaswamy et al., 2013b). We gathered existing data quality dimensions in table 2.

Table 2: IoT data quality dimensions

| *Dimensions* | *References* |
|---|---|
| Accuracy | (C Cappiello & F A Schreiber, 2009), (A Klein & W Lehner, 2009), (van der Togt, Bakker, & Jaspers, 2011), (B Guo, D Zhang, Z Wang, Z Yu, & X Zhou, 2013), (Ramaswamy et al., 2013a), (S Sicari, Rizzardi, Cappiello, & Coen-Por, 2014), (Fan, 2015b), (Lawson & Ramaswamy, 2015), (Karkouch, Al Moatassime, Mohsannif, & Noel, 2015), (Barnaghi, Bermudez-edo, & Tonjes, 2015), (Sabrina Sicari, Rizzardi, Miorandi, Cappiello, & Coen-Porisini, 2016a), (Karkouch et al., 2016b), (V. J. Lawson & Ramaswamy, 2016b), (Qin et al., 2016a), (Cai, Xu, Jiang, & Vasilakos, 2017), (Liono, Jayaraman, Qin, Nguyen, & Salim, 2018), (Perez-Castillo, Carretero, Rodriguez, Caballero, & Piattini, 2018), (Sabrina Sicari, Rizzardi, Cappiello, Miorandi, & Coen-Porisini, 2018), (Wei et al., 2019a), (Côrte-Re, Ruivo, & Oliveira, 2019), (Mavrogiorgou et al., 2019a), (Liono, Jayaraman, Qin, Nguyen, & Salim, 2019), (Anagnostopoulos & Kolomvatsos, 2019), (Zubair et al., 2019), (Côrte-Real et al., 2020), (Byabazaire, O'Hare, & Delaney, 2020) |
| Precision | (Aggarwal, Ashish, & Sheth, 2013), (Zubair et al., 2019), (Anagnostopoulos & Kolomvatsos, 2016), |
| Usability | (Barnaghi, Bermudez-Edo, & Tönjes, 2015) |
| Relevance | (Zubair et al., 2019), (Qin et al., 2016b), (Perez-Castillo, Carretero, Rodriguez, Caballero, Piattini, et al., 2018) |
| Ease of Understanding | (Islam, Kwak, Kabir, Hossain, & Kwak, 2015) |
| Privacy | (Zubair et al., 2019), (Liono et al., 2019), (Setia et al., 2013), (Côrte-Real et al., 2020), (Cinzia Cappiello & Fabio A Schreiber, 2009), (Anja Klein & Wolfgang Lehner, 2009b), (Ramaswamy et al., 2013b), (V. Lawson & Ramaswamy, 2015), (Fan, 2015a), (Karkouch et al., 2016a), (Sabrina Sicari, Rizzardi, Coen-Porisini, & Cappiello, 2014), (Sabrina Sicari, Rizzardi, et al., 2016b), (Wei et al., 2019b), (Luo, Huang, Kanhere, Zhang, & Das, 2019) |
| Objectivity | (Čolaković & Hadžialić, 2018) |
| Reputation | (Sabrina Sicari, Rizzardi, et al., 2016b) |
| Granularity | (Barnaghi, Bermudez-Edo, & Tönjes, 2015) |
| Integrity | (Čolaković & Hadžialić, 2018) |
| Currency | (Zubair et al., 2019), (Byabazaire et al., 2020) |
| Completeness | (C Cappiello & F A Schreiber, 2009), (A Klein & W Lehner, 2009), (van der Togt et al., 2011), (Aggarwal, Ashish, & Sheth 2013), (S Sicari et al., 2014), (Fan, 2015b), (Karkouch et al., 2016b), (Sabrina Sicari, Rizzardi, et al., 2016a), (Liono et al., 2018), (Sabrina Sicari et al., 2018), (Wei et al., 2019a), (Côrte-Re et al., 2019), (Mavrogiorgou et al., 2019a), (Liono et al., 2019), (Anagnostopoulos & Kolomvatsos, 2019), (Zubair et al., 2019), (Côrte-Real et al., 2020), (Byabazaire et al., 2020) |
| Timeliness | (C Cappiello & F A Schreiber, 2009), (A Klein & W Lehner, 2009), (Ramaswamy et al., 2013a), (Lawson & Ramaswamy, 2015), (Fan, 2015b), (Karkouch et al., 2016b), (S Sicari et al., 2014), (Sabrina Sicari, Rizzardi, et al., 2016a), (V. J. Lawson & Ramaswamy, 2016b), (Sabrina Sicari, Cappiello, et al., 2016), (Liono et al., 2018), (Sabrina Sicari et al., 2018), (Côrte-Re et al., 2019), (Wei et al., 2019a), (Luo, Kanhere, Zhang, & Das, 2019), (Anagnostopoulos & Kolomvatsos, 2019), (Zubair et al., 2019), (Liono et al., 2019), (Byabazaire et al., 2020) |
| Trustworthiness | (Anagnostopoulos & Kolomvatsos, 2016), (Zubair et al., 2019), (Côrte-Real et al., 2020), (Byabazaire et al., 2020), (Ramaswamy et al., 2013b), (Bin Guo et al., 2013), (V. Lawson & Ramaswamy, 2015), (Karkouch et al., 2016a), (Karkouch et al., 2018), (Barnaghi, Bermudez-Edo, & Tönjes, 2015) |
| Availability | (Sabrina Sicari, Rizzardi, et al., 2016b), (V. J. Lawson & Ramaswamy, 2016a), (Luo, Huang, et al., 2019), (Zubair et al., 2019), (Ramaswamy et al., 2013b), (Li, Nastic, & Dustdar, 2012) |
| Security | (Karkouch et al., 2016a), (Liono et al., 2019), (Sabrina Sicari et al., 2018), (Mavrogiorgou et al., 2019b), (Jing, Vasilakos, Wan, Lu, & Qiu, 2014) |
| Validity (freshness) | (V. Lawson & Ramaswamy, 2015), (Ramaswamy et al., 2013b), (Sabrina Sicari, Cappiello, et al., 2016), (Perez-Castillo, Carretero, Rodriguez, Caballero, Piattini, et al., 2018), (Luo, Huang, et al., 2019), (Byabazaire et al., 2020), (Li et al., 2012) |
| Frequency (temporal resolution) | (Ramaswamy et al., 2013b), (V. Lawson & Ramaswamy, 2015), (V. J. Lawson & Ramaswamy, 2016a) |
| Confidence | (Karkouch et al., 2016a), (Anagnostopoulos & Kolomvatsos, 2016), |
| Volume (Throughput) | (Klein and Lehner 2009), (Karkouch, Mousannif, et al. 2016), (Anagnostopoulos and Kolomvatsos 2019), (Wei, et al. 2019) |
| Ease of access | (Karkouch et al., 2016a) |
| Interpretability | (Karkouch et al., 2016a) |
| Capacity | (Wei et al., 2019b) |
| Format | (Côrte-Real et al., 2020) |

### 4.2. IoT data quality issues

In the context of the Internet of Things, data are vulnerable to numerous issues that affect their quality. The Internet of Things is one of the most emerging technologies globally and its usage in diverse applications. As another of our contribution, we recognized IoT data quality issues. We gathered existing data quality issues in table 3.

**Duplicate** assesses whether a reading is not collected further or a duplication rate in the patients' databases (Liono et al., 2019; Zubair et al., 2019). Data values can be **faulty** from defective sensors or real outliers. For instance, an exponential increase in a smart house's temperature can show a fire crisis (Liono et al., 2019). **Errors** occur owing to sensor defeats generated by severe environments, age, or network data exploitation. Errors may also happen due to the sensor's faulty installation or utilization outside the various credible operations (Karkouch et al., 2016a). The error indicates making any mistake in capturing or processing the stream of data. For example, connecting data in the different year (Chandola, Banerjee, & Kumar, 2009; Pinto-Valverde, Pérez-Guardado, Gomez-Martinez, Corrales-Estrada, & Lavariega-Jarquín, 2013). Sometimes, related information might be vague to an application owing to a **lack of context**. It can be solved partially by employing semantic information about the data. Nevertheless, the contextual data is often unstructured, hence more challenging to interpret automatically (Zubair et al., 2019). Another viewpoint arises from the mismatch between the provider and user of data (Zubair et al., 2019). When the data value displays any random deviation from the expected range of the given data point, the **outlier** error likely happened (Ramaswamy et al., 2013b). These kinds of issues are hard to find. Although they may be false value, they may be rare (Janssen, van der Voort, & Wahyudi, 2017), which contains invaluable information about the application (Simpson et al., 2017). Precision is related to any noise in converting measures to each other; for instance, the voltage to conversion to quantities such as temperature (Kamilaris, Kartakoullis, & Prenafeta-Boldú, 2017). **Inconsistency** is also popular in IoT data (Qin et al., 2016b), and it occurs when multiple sensors measure the same phenomena whereas they depict different results (Gil, Ferrández, Mora-Mora, & Peral, 2016). **Redundancy** happens when several readings contain the same data point; due to either several readings of one sensor or some readings of multiple sensors (Gil et al., 2016; Qin et al., 2016b). The **uncertainty** can refer to readings of the non-existing devices (Qin et al., 2016b). In some situations, the reading might be considered as different measures by different applications; this requirement may lead to **ambiguity** in the interpretation of data (Gil et al., 2016; Qin et al., 2016b). **Response time** is the average time to process the readings entirely (Wei et al., 2019b). It is mostly caused by network latency among sensors and applications, plus the amount of processing to turn sensor readings to the appropriate format quickly interpreting by dashboards.

Table 3: IoT data quality issues

| Issues | References |
|---|---|
| Duplicate | (Karkouch et al., 2016a), (Aggarwal, Ashish, et al., 2013), (Zhang, Meratnia, & Havinga, 2010), (Perez-Castillo, Carretero, Rodriguez, Caballero, Piattini, et al., 2018) |
| Noisy data | (Chandola et al., 2009), (Pinto-Valverde et al., 2013) |
| Errors | (Karkouch et al., 2016a), (Zubair et al., 2019), (Aggarwal, Ashish, et al., 2013), (V. J. Lawson & Ramaswamy, 2016b) |
| Lake of Context | (V. J. Lawson & Ramaswamy, 2016b), (Perez-Castillo, Carretero, Rodriguez, Caballero, Piattini, et al., 2018), |
| Outliers | (Karkouch et al., 2016a), (Zubair et al., 2019), (Zhang et al., 2010), |
| Response time | (Wei et al., 2019b) |
| Inconsistency | (Karkouch et al., 2016a), (Aggarwal, Ashish, et al., 2013), (Yang, 2017), (Perez-Castillo, Carretero, Rodriguez, Caballero, Piattini, et al., 2018), (Gil et al., 2016), (Vilenski, Bak, & Rosenblatt, 2019), (Qin et al., 2016b) |
| Redundancy | (Qin et al., 2016b), (Gil et al., 2016), (Karkouch et al., 2016a), (Aggarwal, Ashish, et al., 2013) |
| Uncertainty | (Gil et al., 2016), (Qin et al., 2016b), (Aggarwal, Ashish, et al., 2013), |
| Ambiguity | (Gil et al., 2016), (Qin et al., 2016b) |

## 5. Proposed category

As mentioned in section 2.2, a few data quality categories have been developed in the context of IoT, which we analyzed existing categories. In section four, we reviewed existing IoT data quality dimensions and issues. Karkouch et al. (2016a) introduced a category into four categories comprises intrinsic, contextual, representation, and accessibility. They have focused on nine data quality dimensions, while a wide range of dimensions has been identified. Lee et al. (2002) developed a category that includes four classifications. Although there is a similarity between Lee's category and Karkouch's category regarding the number and kind of classifications, Lee's category includes 15 data quality dimensions. His category entails new dimensions compared with Karkouch's category. Zubair et al. have concerned 14 data quality dimensions to develop a category into seven classifications. Concerning existing categories, it can be concluded that data quality dimensions and issues are separated into no category. Moreover, all of them have focused on a few dimensions and issues while there are several different data quality dimensions and issues that would be valuable to pay attention to them. Regarding all dimensions, issues, and categories in IoT, we conclude that it is vital to develop a new category and identify related dimensions and issues in each category. Hence, to better understand IoT data quality dimensions and issues, we expand existing categories and propose a new category into five categories in which data quality dimensions and issues are separated. The proposed category includes 19 IoT data quality dimensions and 13 IoT data quality issues (table 4).

In each category, relevant dimensions enhance IoT data quality while decreasing data quality is rooted in related issues. Therefore, it is essential to separate them from each other while in existing categories, separating data quality dimensions and issues were ignored.

Table 4: Proposed categorization of IoT data quality categories, dimensions, and issues

| IoT Data quality categories | IoT Data quality dimensions | IoT Data quality issues |
|---|---|---|
| Accuracy | Objectivity, Precision | Noisy, Error, Outlier |
| Confidence | Relevance, Ease of understanding, Interpretability, Format, Granularity | Uncertainty, Ambiguity, Lack of context |
| Trustworthiness | Reputation, Privacy, Security, Integrity | Insecurity, Source ambiguity, Inconsistency |
| Timeliness | Validity, Currency | High response time |
| Completeness | Usability, Ease of access, Availability, Throughput, Capacity, Frequency | Duplicate, Redundancy, Incompleteness |

## 6. Potential solutions to IoT data quality issues

IoT relevant issues need to be addressed from the data quality aspect. Some problems arise from IoT infrastructure, and they are created due to the essence of data. As our final contribution, we recognized some potential solutions for security issues from our review results.

### 6.1. Using Artificial Intelligence to overcome security issue

Security and privacy issues are grown owing to a considerable number of devices as well as a shortage of standardized surveys concerning IoT security. Thus, these kinds of studies should be concerned. The privacy and security issues exist in IoT infrastructure, which must be addressed to build trust between users. Addressing some issues such as privacy and security in IoT is already addressed in some prior research (Mohanta, Jena, Satapathy, & Patnaik, 2020; Wei et al., 2019b).

Artificial Intelligence (AI) plays a crucial role in IoT since AI can crunch data successfully to empower us to collect invaluable insights. Machine Learning is a sub-branch of AI and has enormous potential to distinguish the irregularities and patterns in smart sensors' data.

The application of IoT is numerous. Some of the application requires a decision which should be taken before the real event happens. For instance, anticipating the fire in a kitchen and alarm the sound to halt the fire. If machine learning technologies are used in IoT applications, it can be possible. Furthermore, to create the system tamper-proof, it requires to address in IoT system. A practical framework is needed to process and measure tremendous data utilizing machine learning techniques (Mohanta et al., 2020). A literature contribution (Lobato, Lopez, & Duarte, 2016) has reviewed the existing security approach from security information and event management to handling data gathering and processing. They proposed an architecture to detect threats using machine learning in real-time when many false alarms create. Hossain et al. (2019) reviewed security issues in terms of applying machine learning in a smart application. Moreover, the volume of the data generated by devices is vast because the number of devices connected to the network is enormous. Processing and performing computation create difficulties in an IoT environment. Therefore, artificial intelligence comes as a release with other rising technologies to solve IoT's security issue. In fact, IoT and AI can link to improve system analysis, improve operational effectiveness, and improve the precision rate (Mohanta et al., 2020). Ghosh et al. (2018) described that AI could help IoT tremendous volume, heterogeneous data, and unstructured data to calculate real-time and make the system realistic.

Data outliers also are one of the critical indications of data uncertainty. Outliers relate to the class of unreliable readings or those readings which are out of bounds without any particular reason. Outliers or anomaly detection as a class of machine learning techniques improve datasets' quality by making them more consistent. Furthermore, the results' accuracy and reliability are increased because outlier detection represents the first state for unreliable reading (Karkouch et al., 2016a).

## 6.2. Using Blockchain to overcome security issue

Blockchain was introduced with Bitcoin (Chen & Lien, 2014) to solve the double-spending problem. A blockchain is built utilizing cryptography. Its cryptographic hash identifies each block, and each block indicates the hash of the previous block. This creates a link among the blocks, forming a blockchain (Mohanta et al., 2020).

Blockchain technology is a distributed/decentralized network where each of them is connected to others. A block comprises various real transactions and their associated qualities (Wiki, 2017). A new set of applications appear because Blockchain proposes a safe value exchange among entities in the network. To empower IoT devices to conserve security and privacy, the authors (Casado-Vara, de la Prieta, Prieto, & Corchado, 2018) have suggested a Blockchain connected Gateway for Bluetooth Low Energy (BLE). Personal user privacy is preserved as the gateway restricts users' sensitive data from being reached without their permission.

The second generation of Blockchain-enabled with smart contracts has introduced new opportunities for data quality improvements. Concerning data quality in the IoT device layer, Cha et al. (2018) proposed that since theoretically, the Blockchain information is an exact representation of the events that occurred in the real world, the data integrity and quality increases with the adoption of blockchain technology. In fact, the adoption of Blockchain offers an automated means for creating, processing, storing, and sharing information in machine-to-machine communication. Utterly, smart contracts can reduce human errors and improve the accuracy, completeness, and accessibility of data through automating the data creation and storage (Kar, Kasimsetty, Barlow, & Rao, 2019; Mingxiao, Xiaofeng, Zhe, Xiangwei, & Qijun, 2017). On top of this, Blockchain stored data in blocks are cryptographically sealed.

Meanwhile, stored data is irreversible and cannot be altered arbitrarily, denoting a high data security degree. In this line, Azaria et al. (2016) proposed a Blockchain-based medical record management system that improved data quality. Casado et al. (2018) proposed a Blockchain-based system to assess and enhance sensor data quality in an IoT platform. DoS attacks are also not a severe concern in Blockchain-based IoT platforms due to the distributed structure of Blockchain. The Blockchain has no single point of failure problem in this case (Bao, Shi, He, & Chood, 2018) in the application layer; there is a threat of identity forgery attacks. An external attacker may try to use a fabricated signature of a node to forge its identity. Similarly, due to the use of a secure digital signature algorithm of the Blockchain, such an attack will not work. Therefore, it can be argued that Blockchain has enhanced different aspects of data quality in general and IoT in particular (Bao et al., 2018).

## 7. Conclusions

Nowadays, the Internet of Things (IoT) is one of the most swiftly utilized technologies in multiple applications. Data is inspiring agriculture, healthy, manufacturing, and diverse business decisions. It is crucial to consider and evaluate the quality of data since poor choices are rooted in low data quality.

In this article, we reviewed data quality in the context of IoT. We reviewed data quality dimensions (section 3) and data quality issues (section 4) in both general and specific domains. Moreover, we reviewed existing data quality categories. It is important to note that current IoT data categories are not many. Concerning existing IoT data quality dimensions, issues, and categories, we proposed a new category and separate IoT data quality dimensions and IoT data quality issues to the best of our knowledge (section 5). This category would enormously aid the practitioners, researchers, and industry people to recognize related dimensions and issues in each category to enhance the quality of data and decline issues in the context of IoT.

Data are vulnerable to manifold problems. In response to another gap in the IoT literature in data quality, we discuss various issues, identified new issues, and their potential solutions (section 6). We found that machine learning, Artificial intelligence, and Blockchain technology have already been recognized to overcome security and privacy issues from the review.

The most critical limitation we encountered was the low number of IoT data quality categories. In fact, categories in the IoT context were not many. Thus, we developed our proposed category based on existing dimensions, issues, and three types we explained in section 3. Moreover, there are no distinct differences between dimensions and issues in the literature, and we try to develop a new category based on our knowledge. However, there are some differences between our proposed category and existing categories.